\documentclass{ws-procs961x669} 

\def\be{\begin{equation}}
\def\ee{\end{equation}}

\usepackage{array}
\usepackage{multirow}
\usepackage{mdwlist}
\usepackage{amsmath}
\usepackage{amssymb}

\begin{document}
\title{\Large Warp drive dynamic solutions\\considering different
      fluid sources}
\author{Osvaldo L.\ Santos-Pereira$^{1,*}$, Everton M.\ C.\
	Abreu$^{2,3,4,\dag}$ and Marcelo B.\ Ribeiro$^{1,4,\S}$}
\address{$^1$Physics Institute, Universidade Federal do Rio de
	Janeiro, Rio de Janeiro, Brazil,\\
	$^2$Physics Department, Universidade Federal Rural do
	Rio de Janeiro, Serop\'edica, Brazil, \\
	$^3$Physics Department, Universidade Federal de Juiz de
	Fora, Juiz de Fora, Brazil\\ $^4$Applied Physics Graduate
	Program, Physics Institute, Universidade Federal do Rio de
	Janeiro, Rio de Janeiro, Brazil\\
E-mails: $^*$olsp@if.ufrj.br,
         $^{\dag}$evertonabreu@ufrrj.br,
         $^{\S}$mbr@if.ufrj.br}
\begin{abstract}
Alcubierre proposed in 1994 that the well known special relativistic
limitation that particles cannot travel with velocities bigger than the
light speed can be bypassed when such trips are considered globally within
specific general relativistic frameworks. Although initial results indicated
this scenario as being unphysical, since it would seem to require negative
mass-energy density, recent theoretical analyses suggest that such an unphysical 
situation may not always be necessarily true. In this paper, we present solutions 
of the Einstein equations using the original Alcubierre warp drive metric 
endowed with various matter-energy sources, namely dust, perfect fluid, 
anisotropic fluid, and perfect fluid within a cosmological constant spacetime. 
A connection of some of these solutions featuring shock waves described by the 
Burgers equation is also shown.
\end{abstract}

\bodymatter

\section{Introduction}

Alcubierre\cite{Alcubierre1994} advanced a model based on a specific general
relativistic spacetime geometry in which massive particles can travel at
superluminal speeds if they are located inside a specially designed spacetime
distortion. The physics of this possible propulsion method, named as
\textit{warp drive} (WD) after science fiction literature, consists of a
special spacetime metric forming a spacetime distortion, called
\textit{warp bubble}, such that it generates an expansion behind the
distortion and a contraction in front of it. That makes it possible for a
massive particle to be in a local sub-luminal speed inside the bubble, as
required by special relativity, whereas outside the particle is propelled
at superluminal speeds.  Therefore, employing of this spacetime distortion
a massive particle travels between two points in spacetime with apparent
time less than the time a light particle with zero mass would require to
travel between the same two points.

The WD metric, as originally proposed by Alcubierre, is based on a general
metric that uses the 3+1 formalism.\cite{adm1,adm2} It consists of a boost
in the direction of one of the spatial coordinates, described by the shift
vector, a function of spacetime coordinates and the product of two functions,
the velocity of the center of the warp bubble and the shape function that
controls the shape of the bubble. Although Alcubierre did not solve the
Einstein equations with his proposed spacetime geometry, he gave an example
of a shape function in the form of a hat function.\cite{Alcubierre1994} One
of the main caveats of the Alcubierre WD spacetime is the requirement of
negative mass-energy density and the violation of the dominant and weak
energy conditions.
 
Ford and Roman\cite{FordRoman1996} derived quantum inequalities for free
massless quantized scalar fields, electromagnetic fields, and massive
scalar fields in four-dimensional Minkowski spacetime and concluded that
these constraints have the form of an uncertainty principle limitation on
the magnitude and duration of negative mass-energy densities and that the
exotic solutions of Einstein equations such as wormholes and WD would have
significant limitations in their viability. Pfenning and
Ford\cite{Pfenning1997} followed this last work and calculated the upper
bound limits necessary for the WD viability, concluding then that the energy
required to create a warp bubble is ten orders of magnitude greater than
the total mass of the entire visible universe, also negative.

Krasnikov\cite{Krasnikov1998} approached the hyperfast interstellar travel
problem in general relativity by discussing the possibility of whether
or not a mass particle can reach a remote point in spacetime and return
sooner than a photon would. Krasnikov argued that it is not possible for a
mass particle to win this race under reasonable assumptions for globally
hyperbolic spacetimes. He discussed in detail the specific spacetime
topologies, but with the constraint that tachyons would be a requirement
for superluminal travel to occur. He conjectured a spacetime modification
device that can be used to make superluminal travel possible without
tachyons. Such spacetime was named as \textit{Krasnikov tube} by Everet
and Roman,\cite{EveretRoman1997} and they generalized the metric proposed
by Krasnikov by hypothesizing a tube along the path of the particle connecting
Earth to a distant star.

The Krasnikov metric cannot shorten the time for a one-way trip from Earth
to a distant star, but it can make the time for a round trip arbitrarily
short for clocks on Earth. However, the Everet and Roman\cite{EveretRoman1997}
extension of the Krasnikov metric has the property that inside the tube
the spacetime is flat and the lightcones are opened in such a way that
they allow the superluminal travel in one direction. They also mentioned
that even though the Krasnikov tube does not involve closed timelike curves
it is possible to construct a time machine with a system of two
non-overlapping tubes, and demonstrated that the Krasnikov tube also
requires thin layers of negative energy density and large total negative
energies.

Lobo and Crawford\cite{Lobo2002} discussed the Krasnikov metric in detail,
addressed the violation of the weak energy condition and the viability
of superluminal travel. These authors concluded that with the imposition of
the weak energy condition the Olum theorem\cite{Olum1998} prohibits
superluminal travel and pointed out the necessity of further research on
spacetimes with closed timelike curves and the need for a precise definition
of superluminal travel. They argued that one can construct metrics that
allow superluminal travel but are flat Minkowski spacetimes.  
The quantum inequalities, brought from quantum field theory,
\cite{EveretRoman1997} were also discussed. Van de Broeck\cite{Broeck1999} 
showed how a minor modification of the Alcubierre WD geometry can reduce the 
total energy required for the creation of a warp bubble. He presented a 
modification of the original WD metric where the total negative mass would 
be of the order of just a few solar masses. 

Nat\'ario \cite{Natario2002} proposed a new version of the WD theory with
zero expansion and questioned the effects that occur in the WD in
opposition to his newly proposed metric, namely, the zero expansion
Nat\'ario WD metric. Nat\'ario discussed the nature of the WD spacetime
symmetries with a series of propositions and corollaries such as the proof
that the WD spacetime is flat whenever the tangent vector to the Cauchy
surfaces is a Killing vector field for the Euclidean metric despite
being time-dependent and as a particular case. In addition, it is also
flat wherever the tangent vector is spatially constant. He also showed 
that nonflat WD spacetime violate both the \textit{weak energy condition}
(WEC) and the \textit{strong energy condition} (SEC), as well as that the
WD metric can be obtained from the Nat\'ario metric by a particular choice
of coordinates. The spherical coordinates were the choice of charts for the
zero expansion of the Nat\'ario WD spacetime.

Lobo and Visser\cite{LoboVisser2004} pointed out that the WD theory is an
example of reverse engineering of the solutions of the Einstein equations
where one defines a specific spacetime metric and then one finds the matter
distribution responsible for the respective geometry. These authors verified
that the class of WD spacetimes necessarily violate the classical energy
conditions even for low warp bubble velocity. Hence, this is the case of
a geometric choice and not of superluminal properties. They proposed a more
realistic WD by applying linearized gravity to the weak WD with nonrelativistic
warp bubble velocities and argued that for the Alcubierre WD and its version
proposed by Nat\'ario\cite{Natario2002} the center of the bubble must be
massless. They found that even for low velocities the negative energy stored
in the warp fields must be a significant fraction of the particle's mass at
the center of the warp bubble. 

White\cite{White2003,White2011} described how a warp field interferometer
could be implemented at the Advanced Propulsion Physics Laboratory with the
help of the original\cite{Alcubierre1994} WD ideias. He also pointed out that
the expansion behind the warp bubble and contraction in front of it is due
to the nature of the WD functions, and that the distortion of spacetime may
be interpreted as a kind of Doppler effect or stress and strain on spacetime.
Lee and Cleaver\cite{LeeCleaver2016} argued that external radiation might
affect the WD, and that the warp field interferometer proposed by
White\cite{White2003,White2011} could not detect spacetime distortions. 
Mattingly et al.\cite{Mattingly2020} discussed curvature invariants in the
Nat\'ario WD.

Bobrick and Martire\cite{Bobrick2021} proposed a general WD spacetime that
encloses all WD definitions removing any alleged issues with the original
Alcubierre WD. They presented a general subluminal model with spherical
symmetry and positive energy solutions that satisfies the quantum energy
inequalities such that it reduces two orders of magnitude the WD requirement
for negative energy density. They claimed that any type of WD, including
the original one, is a place of regular or exotic material moving in
inertial form with a certain speed.

Lentz\cite{Lentz} claimed that the original warp bubble proposed by
Alcubierre\cite{Alcubierre1994} can be physically interpreted as hyper-fast
gravitational solitons and presented the first solution for superluminal
solitons in general relativity satisfying the WEC and the momentum conditions
for conventional sources of stress, like energy and momentum, that do not
require large amounts of negative energy. The basis for his work is to
assume that the shift vector components obey a kind of wave equation giving
rise to a positive energy geometry. Fell and Heisenberg\cite{Fell2021} also
addressed the WD as gravitational solitons and shed light on the Eulerian
energies and their relation to the WEC, raising the possibility for
superluminal spacetimes with viable amounts of energy density. 

Quarra\cite{Quarra2021} established that within the scope of general
relativity certain gravitational waveforms can result in geodesics that
arrive at distant points earlier than light signals in flat spacetime,
and presented an example waveform that can be used to manifest superluminal
behavior.

Santiago et {al.}\cite{Santiago2021a} claimed that generic WD metrics
violate the \textit{null energy conditions} (NEC) and discussed how Eulerian
observers are privileged, meaning that these observers may perceive positive
energy densities causing the impression of viable WD. They also argue that
for the WD to become a possibility it would require that all timelike
observers observe positive energy densities. They stated that any WD spacetime
will unavoidably violate the energy conditions. In a subsequent
work\cite{Santiago2021b} they claimed that exotic spacetimes, such as the
WD one will always violate energy conditions. They also provided other
examples of such spacetimes relating them to wormholes, tractor beams and
stress beams.  

For a detailed discussion on the basics of WD theory, the reader is referred
to Alcubierre and Lobo.\cite{AlcubierreLobo2017} For a nice exposition on
WD theory one is referred to the lecture notes of the course given by
Shoshany.\cite{Shoshany2019} It is particularly noteworthy the recent results
concerning the possibility of WD features in negative energy density
distribution of an experimental Casimir cavity,\cite{casimir} results that 
open the striking possibility of ingenious experimental avenues on how to
create the WD phenomena in a laboratory environment. 

In this work, we present a summary of the results we have recently obtained
when solving the Einstein equations with the assumption of simple fluid
distributions embedded in the original Alcubierre WD spacetime geometry.
New results such as the vacuum solutions connecting the WD to shock waves
via the Burgers equation are presented.\cite{nos1} We used the perfect fluid
energy-momentum tensor\cite{nos2} (EMT) which disclosed new exact solutions
of the Einstein equations. We also demonstrated that starting with simple
forms of mass and energy sources, vacuum solutions for the WD spacetime
arise when we impose that the WEC is trivially satisfied by a choice
of gauge on the shift vector and the spacetime coordinates.\cite{nos2}
Further results of the Alcubierre WD metric endowed with a charged dust EMT
that considers an electromagnetic tensor in curved spacetime with the
cosmological constant are mentioned.\cite{nos3} Finally, adding the
cosmological constant to the Einstein equations coupled with the Alcubierre
WD spacetime having perfect fluid as a source has the effect of leading the
energy density to possibly becoming positive depending on both the value
and sign of the cosmological constant.\cite{nos4}

The paper is organized as follows. Section 2 reviews very brief the Alcubierre
WD metric. Section 3 depicts the perfect and anisotropic fluid cases\cite{nos2}
and Section 4 discusses the relationship between the WD spacetime, vacuum
solutions, and shock waves described by the Burgers equations.\cite{nos1} 
Section 5 presents a discussion on the role of the cosmological constant in the 
Einstein equation solutions for the WD spacetime.\cite{nos4} Section 6 presents 
the conclusions.

\section{The Alcubierre Warp Drive Spacetime Metric}

Alcubierre originally used\cite{Alcubierre1994} the Arnowitt-Deser-Misner
(ADM) approach of general relativity,\cite{adm1,adm2} a formalism where the
spacetime is described by a foliation of space-like hypersurfaces of constant 
time coordinate. The general form of the WD metric in this formalism is
described by the following equation,
\begin{eqnarray}
{ds}^2 = -\,d\tau^2 &=& g_{\alpha\beta} dx^{\alpha} dx^{\beta} \nonumber \\
&=& - \Big(\alpha^2 -\beta_i\beta^{i}\Big) \, dt^2 
+ 2 \beta_i \, dx^i \, dt + \gamma_{ij} \, dx^i \, dx^j \,\,,
\label{awdmetric1}
\end{eqnarray}
where $d\tau$ is the lapse of proper time, $\alpha$ is the lapse function,
$\beta^i$ is the space-like shift vector, and $\gamma_{ij}$ is the spatial 
metric for the hypersurfaces. The Greek indices range from 0 to 3, whereas 
the Latin ones indicate the space-like hypersurfaces and range from 1 to 3. 
The lapse function $\alpha$ and the shift vector $\beta^i$ are functions 
to be determined, $\gamma_{ij}$ is a positive-definite metric on each of 
the space-like hypersurfaces, for all values of time, a feature that makes 
the spacetime globally hyperbolic. The lapse function $d\tau$ is a measure 
of proper time  between two adjacent hypersurfaces by observers moving 
along the normal direction to the hypersurfaces (Eulerian observers).
The shift vector is a tangent vector to the hypersurfaces that relates 
the spatial coordinate systems on different hypersurfaces.

Alcubierre assumed the following ad hoc particular choices for the
parameters of the general ADM metric,\cite{Alcubierre1994}
\begin{align}
\alpha &= 1, 
\\
\beta^1& = - v_s(t)f\big[r_s(t)\big], \label{betax}
\\
\beta^2 &= \beta^3 = 0,
\\
\gamma_{ij} &= \delta_{ij}.
\end{align}
Rewriting Eq.\ \eqref{awdmetric1} with these definitions yields the
\textit{Alcubierre WD spacetime metric} below,
\be
ds^2 = - \left(1 - \beta^2\right)dt^2 - \beta \,dx\,dt 
+ dx^2 + dy^2 + dz^2.
\label{awdmetric2}
\ee


\section{Einstein Equations Solutions for Simple Fluids}

We to find solutions to the Einstein equations with the Alcubierre
WD metric as the underlying geometry by coupling simple matter-energy
distributions, namely the dust of particles, the perfect fluid and the
anisotropic fluid with off-diagonal metric terms to see if those EMTs 
could lead to a viable WD bubble.

Starting with incoherent fluid, or dust, vacuum solutions of the Einstein
equations for the WD metric \eqref{awdmetric2} were recovered, which led to
the connection of the WD with shock waves via Burgers equation.\cite{nos1}
Vacuum solutions mean that all the energy conditions are trivially satisfied,
but the possibility of vacuum shock waves that could physically represent
warp bubbles indicated that other known matter-energy distributions could
result in more complex solutions of Einstein equations. Under this perception
we proposed two other energy-momentum tensors, the perfect and
``parametrized'' fluids, where the latter is, in fact, an anisotropic fluid
with heat flux.\cite{nos2}

The EMT for an anisotropic and dissipative fluid may be written by the
following equation,
\be
 T^{\alpha \beta} = \mu u^\alpha u^\beta + p h^{\alpha \beta} 
+ u^\alpha q^\beta + u^\beta q^\alpha  + \pi^{\alpha \beta},
\ee
where
\be
h_{\alpha \beta} = g_{\alpha \beta} + u_a\,u_b 
\ee
projects tensors onto hypersurfaces orthogonal to $u^\alpha$, $\mu$ is
the matter density, $p$ is the fluid static pressure, $q^\alpha$ is the
heat flux vector and $\pi^{\alpha \beta}$ is the viscous shear tensor.
The world lines of the fluid elements are the integral curves of the
four-velocity vector $u^\alpha$. The heat flux vector and the viscous shear
tensor are transverse to the world lines, that is, 

\be
q_a\,u^a = 0, \:\:\qquad  \:\: \mbox{and}\:\: \:\:\qquad  \pi_{ab}\,u^b = 0\,.
\ee

For simplicity we depicted the EMT of the anisotropic fluid in the
following matrix form,
\be
T_{\alpha \sigma} = 
\begin{pmatrix} 
\mu + \beta^2 p  & \quad \!\!\! - \beta D & \quad 0  & \quad 0  \\ 
- \beta D        & \:\: A         & \quad 0  & \quad 0  \\ 
0                & \:\: 0         & \quad B  & \quad 0  \\ 
0                & \:\: 0         & \quad 0  & \quad C 
\end{pmatrix}\,,
\label{emtanisofluid}
\ee
where $p$, $A$, $B$ and $C$ are anisotropic static pressures, $D$ is the
momentum density parameter that represents the heat flux for this fluid,
and $\mu$ is the particles' matter density.
Notice that if we choose all pressures and momentum density as being equal
to the static pressure $p$ we recover the perfect fluid form shown below, 
\be
T_{\alpha \sigma} = 
\begin{pmatrix} 
\mu + \beta^2 p & \quad \!\!\! - \beta p & \quad 0  & \quad 0  \\ 
- \beta p       & \:\:  p         & \quad 0  & \quad 0  \\ 
0               & \:\:  0         & \quad p  & \quad 0  \\ 
0               & \:\:  0         & \quad 0  & \quad p 
\end{pmatrix}\,.
\label{emtperffluid}
\ee

The EMT for a perfect fluid can be written in terms of the tensor notation
as follows,
\be
T_{\alpha \beta} = \left(\mu + p\right) \, u_\alpha u_\beta
+ p \, g_{\alpha \beta}\,,
\ee
where $\mu$ is the matter density, $p$ is the fluid static pressure, 
$g_{\alpha \beta}$ is the metric tensor and $u_{\alpha}$ is the 
4-velocity of an observer inside the fluid. Perfect fluids have no 
shear stress, rotation, heat conduction or viscosity. To recover the
EMT for dust of particles, we have to choose the
static fluid pressure to be zero, and the matrix form for the dust
is given by the following equation,
\be
T_{\alpha \sigma} = 
\begin{pmatrix} 
\mu & 0 & 0  & 0  \\ 
0   & 0 & 0  & 0  \\ 
0   & 0 & 0  & 0  \\ 
0   & 0 & 0  & 0 
\end{pmatrix},
\label{emtdust}
\ee
which can be written in tensor notation as follows,
\be
T_{\alpha \beta} = \mu \, u_{\alpha} u_{\beta}\,,
\ee
where $\mu$ is the matter density represented by a scalar function of the
spacetime coordinates, and $u^\alpha$ are the observers 4-velocity
components.

Table~1 lists the solutions of the Einstein equations for the anisotropic
fluid defined by Eq.\ \eqref{emtanisofluid},\cite{nos2} 
\begin{table}[ht!]
{Table 1:} WD solutions of the Einstein equations with anisotropic fluid
as source. \vspace{3mm} 
\begin{tabular}{| m{3cm} | m{3cm} | m{5.4cm} |}
\hline 
Case & Conditions & Results \\ 
\hline 
\multirow{2}{*}{$1) \
\displaystyle{\frac{\partial \beta}{\partial z} = 0}$}
& 
$1a) \ \displaystyle{\frac{\partial \beta}{\partial x} = 0}$
&
$\begin{array} {ll} 
\\ [-10pt]
\displaystyle{\mu = \beta^2(2D-A-p) + \frac{A}{3}} \\ [7pt]
\beta = \beta(t,y)\\ [7pt]
\displaystyle{B = - C = \frac{A}{3}}\\ [7pt]
\displaystyle{\left(\frac{\partial \beta}{\partial y}
\right)^2 = 32 \pi C},  \\ [10pt]
\displaystyle{\frac{\partial^2\beta}{\partial y^2} =
16 \pi \beta (A-D)}
\\[7pt]
\end{array}$ \\
\cline{2-3}   
&
$1b) \ \displaystyle{\frac{\partial \beta}{\partial y} = 0}$
&
$\begin{array} {ll} 
\\ [-10pt]
\displaystyle{\mu = - \beta^2p}\\ [7pt]
\beta = \beta(t,x)\\ [7pt]
\displaystyle{B = C}  \\ [5pt]
\displaystyle{A = D = 0} \\ [7pt]
\displaystyle{\frac{\partial}{\partial x}\left[
\frac{\partial \beta}{\partial t} 
+ \frac{1}{2} \frac{\partial}{\partial x}(\beta^2)
\right] = - 64 \pi B} \\[8pt]
\rightarrow\mbox{solution \textit{dismissed} as unphysical} \\
\end{array}$ \\
\hline 

\multirow{2}{*}{$2) \
\displaystyle{\frac{\partial \beta}{\partial y} = 0}$}
& 
$2a) \ \displaystyle{\frac{\partial \beta}{\partial x} = 0}$
&
$\begin{array} {ll} 
\\ [-10pt]
\displaystyle{\mu = \beta^2(2D-A-p) + \frac{A}{3}} \\ [7pt]
\beta = \beta(t,z)\\ [7pt]
\displaystyle{B = - C = \frac{A}{3}}\\ [7pt]
\displaystyle{\left(\frac{\partial \beta}{\partial z}
\right)^2 = 32 \pi C}  \\ [9pt]
\displaystyle{\frac{\partial^2\beta}{\partial z^2} =
16 \pi \beta (A-D)}
\\[7pt]
\end{array}$ \\
\cline{2-3}   
&
$2b) \ \displaystyle{\frac{\partial \beta}{\partial z} = 0}$
&
$\begin{array} {ll} 
\\ [-10pt]
\displaystyle{\mu = - \beta^2p}\\ [7pt]
\beta = \beta(t,x)\\ [7pt]
\displaystyle{B = C}  \\ [5pt]
\displaystyle{A = D = 0} \\ [7pt]
\displaystyle{\frac{\partial}{\partial x}\left[
\frac{\partial \beta}{\partial t} 
+ \frac{1}{2} \frac{\partial}{\partial x}(\beta^2)
\right] = - 64 \pi B}
\\[8pt]
\rightarrow\mbox{solution \textit{dismissed} as unphysical} 
\end{array}$ \\  
\hline 
\end{tabular}
\caption{Summary of all solutions of the Einstein equations for the
	 WD metric having the parametrized perfect
	fluid as source of energy and momentum.}
\label{tableppf}
\end{table}
where the two solutions subsets 1b and 2b are considered unphysical because
the resulting EMT leads to the anisotropic static pressures $A$ and $D$
equal to zero, whereas $B=C$. The energy density in the $T_{00}$
component of the EMT is equal to zero, resulting in the following equation
of state relating the matter density $\mu$, the shift vector $\beta$ and the
static pressure $p$, 
\be
\mu = - \beta^2 p \,.
\ee
We must point out that we made a notational change on the shift vector
definition $\beta^1$ so that it has the negative sign of the original
parameter defined by Alcubierre,\cite{Alcubierre1994} yielding the
shift vector $\beta$ used in here as being given as below, 
\be
\beta = - \beta^1 = v_s(t)f\big[r_s(t)\big]\,.
\label{newshift}
\ee
Finally, one should also notice that solutions 1b and 2b brought about
the following Burgers equation in dissipative form, 
\be
\frac{\partial}{\partial x}\left[
\frac{\partial \beta}{\partial t} 
+ \frac{1}{2} \frac{\partial}{\partial x}(\beta^2)
\right] = - 64 \pi B \,.
\label{burgerspi}
\ee

Perfect fluid solutions have equations of state given by the expression
below,\cite{EllisElst}
\be
p = p(\mu)=(\gamma-1)\mu\,\,, 
\label{eq-state}
\ee
where $\gamma=$ is a constant which for ordinary fluids can be approximated
by $1\leq\gamma\leq 2$. The incoherent matter, or dust, corresponds to
$\gamma = 1$, and radiation corresponds to $\gamma=\frac{4}{3}$. For
solutions 1b and 2b in Table 1 it is clear that $\gamma$ is a function of
the shift vector given by the following equation,
\be
\gamma = 1 - \frac{1}{\beta^2} \,\,.
\label{pfgamma1}
\ee
Notice that since the shift vector is a function of the warp velocity and it
is also the regulating function, we are facing a problem with a discontinuity
when $\beta = 0$ since outside the warp bubble the regulating function
approaches to zero, but inside the warp bubble $f(r_s) = 1$. So Eq.\
\eqref{pfgamma1} takes the following form,
\be
\gamma = 1 - \frac{1}{v_s^2} \,,
\label{pfgamma2}
\ee
where $v_s = v_s(t)$ is the warp bubble velocity. Solutions 1a and 2a shown
in Table 1 are very similar, except that for solution 1a the shift vector is
a function of spacetime coordinates $(t,y)$, whereas solution 2a is a function
of $(t,z)$ coordinates. However, in both of these cases, the shift vector may
be a complex-valued function depending on the values of the static pressure
$C$. For these solutions the equations of state relating the matter density
$\mu$ and the fluid pressures are identical and given by the following
expression,
\be
\mu = \beta^2(2D-A-p) + \frac{A}{3}\,.
\label{eqstate1}
\ee
The perfect fluid solutions listed in Table~2 are a special case of the
anisotropic fluid if we chose all pressures to be equal to $p$.
Table 3 presents the special dust case ($p=0$), which leads to vacuum.
\begin{table}[ht!]
{Table 2:} WD solutions of the Einstein equations with perfect fluid as source.
\vspace{3mm} 
\begin{tabular}{| m{3cm} | m{3cm} | m{5.4cm} |}
\hline 
Case & Condition & Results \\ 
\hline 
\multirow{2}{*}{$1) \
\displaystyle{\frac{\partial \beta}{\partial z} = 0}$}
& 
$1a) \ \displaystyle{\frac{\partial \beta}{\partial x} = 0}$
&
$\begin{array} {ll} 
p = 3 \mu \\ [6pt]
\beta = \beta(y,t)\\ [6pt]
\displaystyle{\frac{\partial \beta}{\partial y} 
=\pm\sqrt{-32 \pi \mu}} \\ [8pt]
\end{array}$ \\ [28pt]
\cline{2-3}   
&
$1b) \ \displaystyle{\frac{\partial \beta}{\partial y} = 0}$
&
$\begin{array} {ll} 
p = 3 \mu = 0 \\ [6pt]
\beta = \beta(x,t)\\ [6pt]
\displaystyle{\frac{\partial \beta}{\partial t} 
+ \frac{1}{2} \frac{\partial}{\partial x}(\beta^2)
= h(t)} \\ [8pt]
\end{array}$ \\ [28pt] 
\hline 

\multirow{2}{*}{$2) \
\displaystyle{\frac{\partial \beta}{\partial y} = 0}$}
& 
$2a) \ \displaystyle{\frac{\partial \beta}{\partial x} = 0}$
&
$\begin{array} {ll} 
p = 3 \mu \\ [6pt]
\beta = \beta(z,t)\\ [6pt]
\displaystyle{\frac{\partial \beta}{\partial z} 
= \pm\,\sqrt{-32 \pi \mu}}\\ [8pt]
\displaystyle{\frac{\partial \beta}{\partial z} 
=\,\pm \sqrt{\pm\,96 \pi \mu}} \\ [8pt]
\end{array}$ \\ [28pt]
\cline{2-3}   
&
$2b) \ \displaystyle{\frac{\partial \beta}{\partial z} = 0}$
&
$\begin{array} {ll} 
p = 3 \mu = 0 \\ [6pt]
\beta = \beta(x,t)\\ [6pt]
\displaystyle{\frac{\partial \beta}{\partial t} 
+ \frac{1}{2} \frac{\partial}{\partial x}(\beta^2)
= h(t)} \\ [8pt]
\end{array}$ \\ [28pt] 
\hline 
\end{tabular}
\label{tablepf}
\end{table}
\begin{table}[ht!]
{Table 3:} WD solutions of the Einstein equations with dust source
(leads to vacuum).  \vspace{3mm} 
\begin{tabular}{| m{3cm} | m{3cm} | m{5.4cm} |}
\hline 
Case & Consequence & Results \\ 
\hline 
$1) \ \displaystyle{\frac{\partial \beta}{\partial z} = 0} $ 
& 
$\displaystyle{\frac{\partial \beta}{\partial y} = 0}$
&
$\begin{array} {ll} 
\mu = 0 \\ [6pt]
\beta = \beta(t,x) \\ [6pt]
\displaystyle{\frac{\partial \beta}{\partial t} 
+ \frac{1}{2} \frac{\partial}{\partial x}(\beta^2)
= h(t)}
\\[10pt]
\end{array}$ \\ 
\hline 
$2) \ \displaystyle{\frac{\partial \beta}{\partial y} = 0}$
&
$\displaystyle{\frac{\partial \beta}{\partial z} = 0}$ 
& 
$\begin{array} {ll} 
\mu = 0 \\ [6pt]
\beta = \beta(t,x)\\ [6pt]
\displaystyle{\frac{\partial \beta}{\partial t} 
+ \frac{1}{2} \frac{\partial}{\partial x}(\beta^2)
= h(t)}
\\[10pt]
\end{array}$ \\
\hline 
\end{tabular}
\caption{Summary of results for the WD spacetime having dust
matter content.}
\label{tabledust}
\end{table}


\section{Warp Drive and Shock Waves}

In previous studies\cite{nos1,nos2} it was found an intrinsic relationship
between the WD concept and shock waves via the Burgers equation arising from
vacuum solutions of Einstein equations for the Alcubierre WD metric. Such a
result may mean that the warp bubble could be interpreted as shock waves in
vacuum. In other words, the Burgers equation can be seen as a vacuum solution
of Einstein equations connecting the Alcubierre WD metric to shock waves.

In the case of the anisotropic fluid, we considered as unphysical the solution
which led to a Burgers equation \eqref{burgerspi} because the parameter
values resulted in an EMT with only two static pressures (diagonal terms)
and no energy density. Nevertheless, for the perfect fluid and dust EMTs we
found the Burgers equation in its viscous form describing a dissipative
system with the right-hand side being a function of time coordinate only,
according to the expression below,
\be
\frac{\partial \beta}{\partial t}
+ \frac{1}{2}\frac{\partial}{\partial x} (\beta^2)
= h(t)\,\,.   
\label{burg}
\ee
Here $h = h(t)$ is an arbitrary function to be determined by the boundary
conditions. If $h(t) = 0$ Eq.\ \eqref{burg} takes its homogeneous form
describing a conservative system known as inviscid Burgers equation.

The Burgers or Bateman-Burgers, equation is a well-known partial
differential equation that models several physical systems such as gas
dynamics, traffic flows, and even stock market, since, for the latter
application, it is connected to the symmetries of the Black-Scholes equation.
Its most notorious system description is, however, the phenomena arising
from conservation laws and formation of shock waves, that is, discontinuities
that appears after a finite time and then propagates in a regularly. The
physics behind the field solutions of the Burgers equation can be seen as a
current density.

The WD is depicted by the shift vector $\beta = v_s(t) f(r_s)$, where $v_s$
is the bubble velocity, $f(r_s)$ is the regulating function of the
warp bubble shape and the inviscid Burgers in Eq.\ \eqref{burg} can be
interpreted and representing a conservation law for this current density.
Analyzing each term of Eq.\,\eqref{burg}, the first one on the l.h.s., 
$\partial \beta/\partial t$ can be interpreted as a force per unit mass,
i.e., the time derivative of momentum. The second term on the l.h.s.,
$\frac{1}{2}\,\partial (\beta^2) /\partial x$, can be interpreted as the
divergence of the total energy, which is entirely kinetic. Physically the
result can be understood by considering the WD metric as conservation of
both energy and momentum in the direction of the wave propagation.
Calculating the divergence of the parametrized, or anisotropic, perfect
fluid energy-momentum tensor, and demanding that it should be zero, one
arrives at the following system,\cite{nos2}
\begin{eqnarray} 
\nonumber
&-& \frac{\partial \beta}{\partial x} (D + \mu) 
- \frac{\partial \mu}{\partial t} 
- \beta \left[\frac{\partial D}{\partial x} 
+ \frac{\partial \mu}{\partial x} 
+ \frac{\partial \beta}{\partial t}(2p + A - 3D)\right] \nonumber \\ 
&& \mbox{} \nonumber \\
&& \qquad \qquad +\;\beta^2\left[\frac{\partial D}{\partial t} 
- \frac{\partial p}{\partial t} 
+ 3\frac{\partial \beta}{\partial x}(D-p)\right]
+ \beta^3 \left(\frac{\partial D}{\partial x} 
- \frac{\partial p}{\partial x}\right)= 0,
\label{divT0}  \\
&& \mbox{} \nonumber \\
&&\frac{\partial A}{\partial x} 
+ \frac{\partial \beta}{\partial t} (D-A)
+ \beta \left[ 3 \frac{\partial \beta}{\partial x} (D-A)
+ \frac{\partial D}{\partial t} 
- \frac{\partial A}{\partial t}\right] \nonumber \\
&& \qquad \qquad \qquad + \;\;\beta^2 \left(\frac{\partial D}{\partial x}
- \frac{\partial A}{\partial x}\right) = 0,
\label{divT1}   \\
&& \mbox{} \nonumber \\
&& \frac{\partial B}{\partial y} +
\beta \frac{\partial \beta}{\partial y} (D-A) = 0,
\label{divT2} \\
&& \mbox{} \nonumber \\
&&\frac{\partial C}{\partial z} +
\beta \frac{\partial \beta}{\partial z} (D-A) = 0.
\label{divT3}
\end{eqnarray}
The perfect fluid zero divergence is recovered by letting 
\be
p = A = B = C = D\,,
\ee
hence, Eqs.\ \eqref{divT0} to \eqref{divT3} become,
\begin{eqnarray}
&-& \frac{\partial \mu}{\partial t} 
- \beta \left(\frac{\partial p}{\partial x}
+ \frac{\partial\mu}{\partial x}\right)= 0,
\label{pf1adivT0} \\
\mbox{} \nonumber \\
&&\frac{\partial p}{\partial x}=\frac{\partial p}{\partial y}=\frac{\partial p}{\partial z}= 0\,\,,
\end{eqnarray}
which means that the static pressure $p$ of the perfect fluid does 
not depend on the spatial coordinates, and Eq.\ \eqref{pf1adivT0} 
reduces to the expression below, 
\be
\frac{\partial \mu}{\partial t} 
+ \beta \frac{\partial \mu}{\partial x} = 0,
\label{divpf0}
\ee
which is a continuity equation, $\mu$ is the fluid density, and $\beta$ is
the flow velocity vector field. Notice that for constant density the
fluid has an incompressible flow. Therefore, all partial derivatives of
$\beta$ in terms of the spatial coordinates vanish, and the flow velocity
vector field has zero divergence, this being a classical fluid dynamics
scenario. The local bubble volume expansion rate is zero, and the WD
metric becomes the Minkowski metric in this scenario. 

If we want to find these results for the dust of particles as a
particular case, we can see that the partial time derivative
of the matter-density is zero and we have
\be
\mu \frac{\partial \beta}{\partial x} = 0,
\label{nulldiv}
\ee
which is immediately satisfied since the matter density is zero for the
dust EMT and the Einstein equations solution for the WD is the vacuum 
one. This is also true for the sets of solutions 1b and 2b
found for the perfect fluid, as shown in Table 2. This result for the
vacuum solutions of the Einstein equations with the WD metric as a
background suggests that the necessary energy to create the associate
shock wave is purely geometrical. It is interesting evidence that the
WD metric can be understood as a spacetime motion equivalent to a shock
wave moving in a fluid, where spacetime itself plays the role of the fluid. 

Considering all these results one can speculate about vacuum energy, 
quantum fluctuations and dark matter as fuel for the feasibility of the
warp bubble, but one has to overcome possible difficulties caused by the
event horizon of the bubble that would require particles with imaginary
mass like tachyons.\cite{AlcubierreMG2021}

\section{The Cosmological Constant}

Bearing in mind the results above led to the next step of adding the
cosmological constant in the Einstein equation to seek solutions
for Alcubierre WD spacetime geometry.\cite{nos4} The motivation came from
Eq.\ (19) in the original Alcubierre work,\cite{Alcubierre1994} which was
obtained by applying the WEC considering Eulerian observers, contracting
the EMT and writing the final expression $G_{\mu\nu} = \kappa T_{\mu\nu}$
in the Einstein equation, whose result reads as follows,
\be
T^{\mu \nu} u_\mu u_\nu = \alpha^2 T^{00} = G^{00} = 
- \frac{v_s^2 \rho^2}{4 r_s^2} \left(\frac{df}{dr_s}\right)^2.
\label{weakeng}
\ee
Here $u^\mu$ and $u_\mu$ are respectively the contravariant and covariant
components of the four-vector velocity of the Eulerian observers given by
\be
u^\mu = \frac{1}{\alpha}(1, \beta^1, \beta^2, \beta^3) \ \ , \ \ 
u_\mu = - (\alpha, 0, 0, 0)\,,
\ee
where $\alpha$ is the lapse time between the constant hypersurfaces of the
ADM-formalism and $\beta^i$ are the three-components of the shift vector
$\pmb{\beta}$. Using the original parameters from the $3+1$-formalism, its
general metric and the chosen chart of coordinates we can write Eq.\
\eqref{weakeng} as below,
\be
T_{\alpha \beta} \, u^\alpha u^\beta 
= - \frac{1}{4} \left[
\left(\frac{\partial \beta}{\partial y} \right)^2  +  
\left(\frac{\partial \beta}{\partial z} \right)^2
\right].
\label{edwd1}
\ee

The above expression is clearly a non-positive term everywhere, 
violating the WEC in a nontrivial way. We then showed that
if we begin with simple energy and momentum sources and impose Eq.\
\eqref{edwd1} to be as identically zero we arrive at vacuum solutions
for the WD spacetime.\cite{nos1,nos2} However, if we include the
cosmological constant in the Einstein equations the expression
\eqref{edwd1} is changed, yielding,
\be
T_{\alpha \beta} \, u^\alpha u^\beta 
= \Lambda - \frac{1}{4} \left[
\left(\frac{\partial \beta}{\partial y} \right)^2  +  
\left(\frac{\partial \beta}{\partial z} \right)^2
\right]\,\,.
\label{edwd2}
\ee
For positive and large enough values of the cosmological constant, it may 
be possible to construct a WD that does not violate the WEC, because the
original negative mass-energy density necessary to create the warp bubble
could become positive.

The solutions for the WD metric which include the cosmological constant
in the Einstein equations\cite{nos4} are presented in Table 4. Cases 1b
and 2b are just the vacuum solutions connecting the WD spacetime to shock
waves via the Burgers equations, and they also require a vanishing
cosmological constant, since both the static fluid pressure $(p)$ and the
matter density $(\mu)$ also vanish. The other two solutions subsets 1a and
2a are also similar to the equally labeled ones in Table 2, but now they
include the cosmological constant and the solutions are no longer required
to be complex if the following conditions are satisfied,
\be
\Lambda - 8\pi \mu \geq 0\,,
\ee
\be
\Lambda - 8\pi p \geq 0\,.
\ee

Taking all these results into account imply that creating a warp bubble
for superluminal travel of massive particles seem to require more complex
forms of matter than the dust for stable solutions, and that the requirement
of negative mass-energy may not be as strict as originally thought for
superluminal spacetime travel with warp speeds. The shift vector in the
direction of the warp bubble movement creates a coupling in Einstein equations
that requires the off-diagonal source terms in the energy-momentum tensor that
represent momentum densities.

Previous authors investigated the WD theory from the geometrical point of
view \cite{Krasnikov1998, EveretRoman1997, Broeck1999, Natario2002,
LoboVisser2004} and came up with other types of WD mechnics, but it now seems
clear that it is also necessary to consider both the energy and momentum
tensors to propose a superluminal propelling system in the WD theory. 
\begin{table}[ht!]
Table 4: WD solutions of the Einstein equations with $\Lambda$ and 
perfect fluid.
\begin{tabular}{| m{3cm} | m{3cm} | m{5.4cm} |}
\hline 
Case & Condition & Results \\ 
\hline 
\multirow{2}{*}{$1) \
\displaystyle{\frac{\partial \beta}{\partial z} = 0}$}
& 
$1a) \ \displaystyle{\frac{\partial \beta}{\partial x} = 0}$
&
$\begin{array} {ll} 
\Lambda =  6 \pi \left(\mu - \frac{p}{3}\right) \\ [6pt]
\beta = \beta(y,t)\\ [6pt]
\displaystyle{\frac{\partial \beta}{\partial y} 
= \pm\sqrt{4 (\Lambda - 8 \pi \mu)}} \\ [8pt]
\displaystyle{\frac{\partial \beta}{\partial y} 
= \pm\sqrt{\frac{4}{3} (\Lambda - 8 \pi p)}} \\ [8pt]
\displaystyle{\beta \frac{\partial \mu}{\partial x} 
+ \frac{\partial \mu}{\partial t} = 0 
\ \ \text{(null divergence)}} \\ [8pt]
\end{array}$ \\ [28pt]
\cline{2-3}   
&
$1b) \ \displaystyle{\frac{\partial \beta}{\partial y} = 0}$
&
$\begin{array} {ll} 
\Lambda = 8 \pi \mu = 8 \pi p = 0\\ [6pt]
\beta = \beta(x,t)\\ [6pt]
\displaystyle{
\frac{\partial \beta}{\partial t}
+ \frac{1}{2} \frac{\partial}{\partial x} 
(\beta^2) = h(t)\, } \\ [6pt]
\text{Null divergence is trivially satisfied}\\ [2pt]
\text{This is the solution found in Ref.\cite{nos1}} \\ [8pt]
\end{array}$ \\ [28pt] 
\hline 

\multirow{2}{*}{$2) \
\displaystyle{\frac{\partial \beta}{\partial y} = 0}$}
& 
$2a) \ \displaystyle{\frac{\partial \beta}{\partial x} = 0}$
&
$\begin{array} {ll} 
\Lambda =  6 \pi \left(\mu - \frac{p}{3}\right) \\ [6pt]
\beta = \beta(y,t)\\ [6pt]
\displaystyle{\frac{\partial \beta}{\partial z} 
= \pm\sqrt{4 (\Lambda - 8 \pi \mu)}} \\ [8pt]
\displaystyle{\frac{\partial \beta}{\partial z} 
= \pm\sqrt{\frac{4}{3} (\Lambda - 8 \pi p)}} \\ [8pt]
\displaystyle{\beta \frac{\partial \mu}{\partial x} 
+ \frac{\partial \mu}{\partial t} = 0 
\ \ \text{(null divergence)}} \\ [8pt]
\end{array}$ \\ [28pt]
\cline{2-3}   
&
$2b) \ \displaystyle{\frac{\partial \beta}{\partial z} = 0}$
&
$\begin{array} {ll} 
\Lambda = 8 \pi \mu = 8 \pi p = 0 \\ [6pt]
\beta = \beta(x,t)\\ [6pt]
\displaystyle{\frac{\partial \beta}{\partial t} 
+ \frac{1}{2} \frac{\partial}{\partial x}(\beta^2)
= h(t)} \\ [6pt]
\text{Null divergence is trivially satisfied}\\ [2pt]
\text{This is the solution found in Ref.\cite{nos1}} \\ [8pt]
\end{array}$ \\ [28pt] 
\hline 
\end{tabular}
\caption{Summary of all solutions of the Einstein equation with
the cosmological constant and the WD metric 
having the perfect fluid EMT as mass-energy 
source. This table is also valid for the dust particle if considered 
as a particular case for the perfect fluid with null pressure $p = 0$.}
\label{tabcosmo}
\end{table}

\section{Conclusions}

In this work, we described new exact solutions of the Einstein equations
endowed with the Alcubierre warp drive (WD) spacetime geometry having simple
energy-momentum tensor (EMT) distributions of matter and energy. The incoherent
matter, or dust, the simplest fluid source EMT used with the WD metric, led
to vacuum solutions and the weak energy condition being satisfied with null
energy density. In addition, we showed that the dynamics of the vacuum WD
spacetime are governed by the Burgers equation, which links the WD bubble to
shock waves.\cite{nos1} Both the perfect fluid EMT and an anisotropic fluid
recover the Burgers equation as vacuum solutions, but some of these new sets of
solutions were considered unphysical, and others showed that the shift vector
can be a complex-valued function.\cite{nos2}
The Einstein equations with the cosmological constant were used with the
WD metric having the perfect fluid as the source, whose results showed that if
the cosmological constant is big enough it would be possible to attain positive
energy density for the WD spacetime.\cite{nos4} This result is in line with
recent research activity in WD theory by other authors, who provide examples
of different WD spacetimes geometries,\cite{Lentz,Fell2021,Bobrick2021} which
suggests that it may be possible to create warp bubbles with positive energy
densities.



\begin{thebibliography}{}

\bibitem{Alcubierre1994}
M.\ Alcubierre, \textit{The warp drive: hyper-fast travel 
within general relativity}. Class.\ Quant.\ Grav. 
11 (1994) L73, arXiv:gr-qc/0009013.

\bibitem{adm1}
R.\ Arnowitt, S.\ Deser and C.W.\ Misner. \textit{Dynamical 
Structure and Definition of Energy in General Relativity}, 
Phys. Rev. 116 (1959) 1322.

\bibitem{adm2}
R.\ Arnowitt, S.\ Deser and C.W.\ Misner. \textit{The 
Dynamics of General Relativity. Gravitation: An introduction 
to current research}, L.\ Witten, (ed.), Wiley, NY, 1962. 
Reprinted in arXiv:gr-qc/0405109.

\bibitem{FordRoman1996}
L.H.\ Ford and T.A.\ Roman, \textit{Quantum Field Theory Constrains 
Traversable Wormhole Geometries}, Phys.\ Rev.\ D 53 (1996) 5496, 
arXiv:gr-qc/9510071.

\bibitem{Pfenning1997}
M.J.\ Pfenning and L.H.\ Ford \textit{The unphysical nature of 
Warp Drive}, Class.\ Quant.\ Grav.\ 14 (1997) 1743, 
arXiv:gr-qc/9702026.

\bibitem{Krasnikov1998}
S.V.\ Krasnikov \textit{Hyperfast Interstellar Travel in General 
Relativity}, Phys.\ Rev.\ D 57 (1998) 4760, arXiv:gr-qc/9511068.

\bibitem{EveretRoman1997}
A.\ Everett and T.A.\ Roman, \textit{A Superluminal 
Subway: The Krasnikov Tube}, Phys.\ Rev.\ D 56 (1997) 2100, 
arXiv:gr-qc/9702049. 

\bibitem{Lobo2002}
F.S.N.\ Lobo and P.\ Crawford, \textit{Weak Energy Condition 
Violation and Superluminal Travel}, Lect.\ Notes Phys.\ 617 
(2003) 277, arXiv:gr-qc/0204038.

\bibitem{Olum1998}
K.\ Olum, \textit{Superluminal travel requires negative energy density}, 
Phys. Rev. Lett, 81 (1998) 3567, arXiv:gr-qc/9805003.

\bibitem{Broeck1999}
C.\ Van Den Broeck,  \textit{A warp drive with more reasonable 
total energy}, Class.\ Quant.\ Grav.\ 16 (1999) 3973, 
arXiv:gr-qc/9905084. 

\bibitem{Natario2002}
J.\ Nat\'ario, \textit{Warp Drive With Zero Expansion},
Class.\ Quant.\ Grav.\ 19 (2002) 1157, arXiv:gr-qc/0110086.

\bibitem{LoboVisser2004}
F.S.N.\ Lobo and M.\ Visser, \textit{Linearized warp drive and the 
energy conditions}, arXiv:gr-qc/0412065.

\bibitem{White2003}
H.G.\ White, \textit{A Discussion of Space-Time Metric 
Engineering}, Gen.\ Relat.\ Grav.\ 35 (2003) 2025. 

\bibitem{White2011}
H.G.\ White, \textit{Warp Field Mechanics 101}, 
J.\ Brit.\ Interplanetary Society 66 (2011) 242.

\bibitem{LeeCleaver2016}
J.\ Lee and G.\ Cleaver, \textit{Effects of External Radiation on an
Alcubierre Warp Bubble}, Physics Essays 29 (2016) 201. 

\bibitem{Mattingly2020}
B.\ Mattingly, A.\ Kar, M.\ Gorban, W.\ Julius, C.\ Watson, M.D.\ Ali,
A.\ Baas, C.\ Elmore, J.\ Lee, B.\ Shakerin, E.\ Davis, and G.\ Cleaver,
\textit{Curvature Invariants for the Accelerating Nat\'ario Warp Drive},
 Particles 3 (2020) 642, arXiv:2008.03366.

\bibitem{Bobrick2021}
A.\ Bobrick and G.\ Martire, \textit{Introducing physical warp drives}, 
Class.\ Quant.\ Grav.\ 38 (2021) 105009, arXiv:2102.06824.

\bibitem{Lentz}
E.W.\ Lentz, \textit{Breaking the warp barrier: Hyper-fast solitons 
in Einstein-Maxwell-plasma Theory}, Class.\ Quant.\ Grav.\ 38 (2021) 
075015, arXiv:2006.07125.

\bibitem{Fell2021}
S.D.B.\ Fell and L.\ Heisenberg, \textit{Positive energy warp drive 
from hidden geometric structures}, Class.\ Quant.\ Grav.\ 38 (2021) 
155020, arXiv:2104.06488.

\bibitem{Quarra2021}
C.J.\ Quarra, \textit{Creating spacetime shortcuts with 
gravitational waveforms}, arXiv:1602.01439.

\bibitem{Santiago2021a}
J.\ Santiago, S.\ Schuster and M.\ Visser, \textit{Generic warp 
drives violate the null energy condition}, arXiv:2105.03079.

\bibitem{Santiago2021b}
J.\ Santiago, S.\ Schuster and M.\ Visser, \textit{Tractor beams, 
pressor beams and stressor beams in General Relativity}, 
Universe 7 (2021) 271, arXiv:2106.05002.

\bibitem{AlcubierreLobo2017}
M.\ Alcubierre and F.S.N.\ Lobo,
\textit{Wormholes, Warp Drives and Energy Conditions},
Fundam.\ Theor.\ Phys.\ 189 (2017) 279, arXiv:2103.05610.

\bibitem{Shoshany2019}
B.\ Shoshany, \textit{Lectures on Faster-than-Light Travel and Time 
Travel}, SciPost Phys.\ Lect.\ Notes, 10 (2019), arXiv:1907.04178.

\bibitem{casimir}  
H.\ White, J.\ Vera, A.\ Han. A.R.\ Bruccoleri and J.\ MacArthur,
\textit{Worldline numerics applied to custom Casimir geometry generates
unanticipated intersection with Alcubierre warp metric},
Eur.\ Phys.\ J.\ C 81 (2021) 677

\bibitem{nos1}
O.L.\ Santos-Pereira, E.M.C.\ Abreu and M.B.\ Ribeiro,
\textit{Dust content content solutions for the Alcubierre warp drive
spacetime}, Eur.\ Phys.\ J.\ C 80 (2020) 786, arXiv:2008.06560.

\bibitem{nos2}
O.L.\ Santos-Pereira, E.M.C.\ Abreu and M.B.\ Ribeiro, 
\textit{Fluid dynamics in the warp drive spacetime geometry}, 
Eur.\ Phys.\ J.\ C\ 81 (2021) 133, arXiv:2101.11467.
 
\bibitem{nos3}
O.L.\ Santos-Pereira, E.M.C.\ Abreu and M.B.\ Ribeiro, 
\textit{Charged dust solutions for the warp drive spacetime},
Gen.\ Relat.\ Gravit.\ 53 (2021) 23, arXiv:2102.05119.

\bibitem{nos4}
O.L.\ Santos-Pereira, E.M.C.\ Abreu and M.B.\ Ribeiro,
\textit{Perfect fluid warp drive solutions with the cosmological 
constant}, Eur.\ Phys.\ J.\ Plus\  136 (2021) 902, arXiv:2108.10960.

\bibitem{EllisElst}
G.F.R.\ Ellis and H.\ van Elst, \textit{Cosmological Models}, 
(Carg\'ese Lectures 1998), NATO Adv. Study Inst.\ Ser.\ C.\ 
Math.\ Phys.\ Sci.\ 541 (1999) 1, arXiv:gr-qc/9812046.

\bibitem{AlcubierreMG2021}
M.\ Alcubierre, talk at the 16th Marcel Grossmann Meeting: AT3-Wormholes,
Energy Conditions and Time Machines (7 July 2021). Available at
\url{https://youtu.be/NqN1c-2fv8Y?t=784}.

\end{thebibliography}
\end{document}